\documentclass[aps,pre,twocolumn,groupedaddress,showpacs,floatfix]{revtex4}
\usepackage{graphicx}
\usepackage{dcolumn}
\usepackage{bm}
\usepackage{amssymb}
\usepackage{amsmath}
\usepackage{epsfig}

\begin{document}
\title{Self-Averaging in the Three Dimensional Site Diluted 
Heisenberg Model at the critical point}
\author{A. Gordillo-Guerrero}  
\affiliation{Departamento de F\'{\i}sica, Universidad de
Extremadura, E-06071 Badajoz, Spain.}
\affiliation{Instituto de Biocomputaci\'on y F\'{\i}sica de Sistemas Complejos (BIFI), Zaragoza, Spain.}
\author{J. J. Ruiz-Lorenzo}
\affiliation{Departamento de F\'{\i}sica, Universidad de
Extremadura, E-06071 Badajoz, Spain.}
\affiliation{Instituto de Biocomputaci\'on y F\'{\i}sica de Sistemas Complejos (BIFI), Zaragoza, Spain.}

\date{May 17, 2007}

\begin{abstract}

  We study the self-averaging properties of the three dimensional site
  diluted Heisenberg model. The Harris criterion~\cite{critharris}
  states that disorder is irrelevant since the specific heat critical
  exponent of the pure model is negative. According with some
  analytical approaches~\cite{harris}, this implies that the
  susceptibility should be self-averaging at the critical temperature
  ($R_\chi=0$).  We have checked this theoretical prediction for a
  large range of dilution (including strong dilution) 
  at critically and we have found that the introduction of
  scaling corrections is crucial in order to obtain self-averageness
  in this model.  Finally we have computed critical exponents and
  cumulants which compare very well with those of the pure model
  supporting the Universality predicted by the Harris criterion.

\end{abstract}

\maketitle
\section{INTRODUCTION}
\label{sec:intro}

It is a well known fact that a possible way to obtain new Universality
classes is to add disorder to pure systems.  The Harris criterion,
Ref.~\cite{critharris}, states that if the specific heat diverges in
the pure system, the disorder will change the critical behavior of the
model, i.e a new Universality class will appear. Conversely, if the
specific heat does not diverge in the pure system the critical
exponents of the disordered system will not change.

The aim of this paper is to study the dependence of some observables
with the disorder (self-averaging properties).  The study of the
self-averaging properties in disordered systems has generated in the
past years a large amount of both analytical, see
Refs.~\cite{harris,korut,harris2}, and numerical, see
Refs.~\cite{wise,derou,sourlas}, works.  We will focus in the
computation at criticality of the quantity $R_\chi$, which will be
defined later and is a measure of the self-averageness of the
susceptibility. 

We will study the three dimensional site diluted Heisenberg model with
quenched disorder (in which, according to Ref.~\cite{critharris}, the
 disorder is irrelevant) in order to test the analytical predictions.

To obtain accurate measures of critical properties, in particular of
$R_\chi$, we will use Finite Size Scaling (FSS) techniques as the
quotient method~\cite{UCMOND3,PERC} that allows us to work in large
lattices at the critical point and to perform infinite volume
extrapolations.

We will show results strongly supporting that this $R_\chi$ cumulant
is zero at the critical point, but only taking into account the
scaling corrections, against some theoretical predictions
(Ref.~\cite{korut}) and supporting others~\cite{harris,harris2}.

The structure of the paper is as follows. In the next section we
summarize some analytical predictions concerning the self-averaging
properties of diluted models at critically. In section~III we define
the model and the observables. In the first part of section~IV we
describe our simulation methods, in the second part we analyze deeply
the correction to scaling exponent, $\omega$, while in the last part
we give our numerical results concerning self-averaging properties of
the susceptibility both in vector and tensor channels.  Results
related to the universality of the critical exponents and cumulants are
given in Appendix~A. Finally we give our conclusions in section~V.

\section{ANALYTICAL PREDICTIONS}
\label{sec:predictions}
The self-averaging (SA) of the susceptibility is defined in terms of:
\begin{equation}
R_\chi \equiv \frac{\overline{ \langle {\cal M}^2 \rangle^2 - 
\overline{\langle {\cal M}^2 \rangle}^2
}}
{\overline{ \langle {\cal M}^2 \rangle}^2 }\quad ,
\label{RsubChi}
\end{equation}
being $\cal M$ the total magnetization. The susceptibility is self-averaging if $R_\chi \to 0$ as $L\to \infty$.

In Ref.~\cite{harris}, the following picture was found:
\begin{enumerate}
\item Outside the critical temperature: $R_\chi=0$. Can be found,
 based in Renormalization Group (RG) or using general statistical
 arguments, that $R_\chi \propto (\xi/L)^d$ in a finite geometry,
 being $L$ the system size and $\xi$ the correlation length, which is
 finite for $T \neq T_c$: then $R_\chi\to 0$ as $L\to \infty$. This is
 called Strong SA.
\item At the critical temperature, a Renormalization Group analysis opens two
  possible scenarios:
\begin{itemize}

\item Models in which according with the Harris criterion the disorder
  is relevant ($\alpha_\mathrm{pure}>0$): $R_\chi\neq 0$. The
  susceptibility at the critical point is not self-averaging. In
  particular, Ref.~\cite{harris} shows that in these conditions
  $R_\chi$ is proportional to the fixed point value of the coupling
  which induces the disorder in the Hamiltonian, which controls the
  new Universality class. This is called no SA.
\item Models in which according with the Harris criterion the disorder
  is not relevant ($\alpha_\mathrm{pure}<0$): $R_\chi=0$. The
  susceptibility at the critical point is self-averaging. In a finite
  geometry $R_\chi$ scales as $L^{\alpha/\nu}\to 0$, where $\alpha$
  and $\nu$ are the critical exponents of the pure system, which are
  the same in the disordered one. This is called Weak SA.
\end{itemize}
\end{enumerate}

The observable $R_\chi$ has been measured in other diluted models: for
example in the four dimensional diluted Ising model, see
Ref.~\cite{ISDIL4D}. In this model a Mean Field computation and a
numerical one found a non zero value for $R_\chi$ although the diluted
model was shown to belong of the same universality class of the pure
model, contradicting the conclusions of Ref.~\cite{harris}.  One can
claim that the logarithms which live in the upper critical dimension 
do the numerical analysis difficult. In particular was found
analytically in Mean Field $R_\chi=0.31024$ and numerically
$R_\chi\in[0.15,0.32]$. Because the logarithms, it was impossible to
do an infinite volume extrapolation for the numerical values of
$R_\chi$.  Notice that in this model the only fixed point is the
Gaussian one (all the values of the couplings are zero) and, following
Ref.~\cite{harris}, $R_\chi$ should be zero.

In addition a two-loops field theory calculation done in
Ref.~\cite{korut} predicts a non zero value for $R_\chi$ for the
diluted Heisenberg model (in which the disorder is irrelevant,
$\alpha_\mathrm{pure}=-0.134$, see Ref.~\cite{peli}). The two-loops
field theoretical prediction for $\alpha$ in the pure case was
$\alpha_\mathrm{pure}>0$, so, apparently, this work is consistent with
the findings of Ref.~\cite{harris}. The starting point in
Ref.~\cite{korut} was the Mean Field computation done in
Ref.~\cite{ISDIL4D}, modifying it to take into account the vector
degrees of freedom, introducing the fluctuations using the
Brezin-Zinn-Justin (BZJ) method, Ref.~\cite{BZJ}. They found
analytically $R_\chi=0.022688$ for the vector channel and universal
(independent of the dilution for all $p<1$). It is important to remark
that in the BZJ method one fixes from the beginning the temperature of
the system to the infinite volume critical one working in a finite
geometry, so in order to compute $R_\chi$ in this scheme the following
sequence of limits is used:
\begin{equation}
R_\chi^*=\lim_{L\to \infty} \lim_{T\to T_c} R_\chi(L,T)\,,
\end{equation}
where $R_\chi^*$ is the infinite volume extrapolation at criticality of
$R_\chi(L,T)$, and $T_c$ is the infinite volume critical temperature 
of the system.
The other possible limits sequence that can be  computed is:
\begin{equation}
\lim_{T\to T_c} \lim_{L\to \infty}  R_\chi(L,T)\,,
\end{equation}
which is zero even when the disorder is relevant since $R_\chi \propto
L^{-d}$ as $T \neq T_c$.

Hence, in order to test these discrepancies we have simulated
numerically the site diluted three dimensional Heisenberg model
computing $R_\chi^*$ in the vector and tensor channels.  To perform
this program, in particular in doing the infinite volume
extrapolations of cumulants and exponents, is really important a proper
use of the corrections to scaling.

\section{THE MODEL}
\label{sec:themodel}

The Heisenberg site-diluted model in three dimensions is defined in terms
of O($3$) spin variables placed in the nodes of a cubic three-dimensional lattice, with
Hamiltonian:
\begin{equation}
H=-\beta\sum_{<i,j>} \epsilon_i \epsilon_j \boldsymbol{\mathit{S}_i}\cdot\boldsymbol{\mathit{S}_j}\, ,
\label{heismodel}
\end{equation}
where the $\boldsymbol{\mathit{S}_i}$ are three-dimensional vectors of
unity modulus, and the sum is extended only over nearest
neighbors. The disorder is introduced by the random variables
$\epsilon_i$ which take on value 1 with probability $p$ and 0 with
probability $1-p$. An actual $\lbrace\epsilon_i\rbrace$ configuration,
will be called a \emph{sample}.

In addition, as done in Ref.~\cite{UCMOND3}, we define a tensorial
channel associated with the vector $\boldsymbol{\mathit{S}}$ through
the traceless tensor:
\begin{equation}
\tau_i^{\alpha\beta}=S_i^\alpha S_i^\beta-\frac{1}{3}\delta^{\alpha\beta}\ ,\qquad
\alpha,\beta=1,2,3\ .
\label{tensorialspin}
\end{equation}

In the following, we shall denote a thermal average with brackets,
while the sample average will be over lined. The observables will be
denoted with calligraphic letters, i.e.  $\cal O$, and with italics the
double average, $O=\overline{\langle\cal O\rangle}$. We define the
total nearest-neighbor energy as:
\begin{equation}
{\cal E} =\sum_{\langle
i,j\rangle}\epsilon_i \epsilon_j \boldsymbol{\mathit{S}_i} \cdot \boldsymbol{\mathit{S}_j}\ ,
\label{totalenergy}
\end{equation}
and the normalized magnetization for both channels as
\begin{equation}
{\cal M}=\frac{1}{V}\sum_i \epsilon_i\boldsymbol{\mathit{S}_i}\ ,
\label{vectorialmag}
\end{equation}
\begin{equation}
{\mathcal M}^{\alpha \beta}_T =\frac{1}{V}\sum_i \epsilon_i(
\mathit{S}^\alpha_i \mathit{S}^\beta_i-\frac{1}{N}\delta^{\alpha\beta}) \,,
\label{tensorialmag}
\end{equation}
being $V$ the volume (defined as $L^3$, where $L$ is the lattice
size).  Because of the finite probability to reach every minimal value
for the free energy, the thermal average of
Eqs.~(\ref{vectorialmag})~and~(\ref{tensorialmag}), is zero in a
finite lattice. Therefore, we have to define the order parameters as
the O($3$) invariant scalars:
\begin{equation}
M=\overline{\left\langle \sqrt{{\cal M}^2}\right\rangle} \quad ,
\quad M_T=\overline{\left\langle \sqrt{\mathrm{tr}
{\mathcal M}_T^2}\right\rangle} .
\label{orderparameters}
\end{equation}

 We also define both susceptibilities as:
\begin{equation}
\chi=V\overline{\left\langle {\cal M}^2 \right\rangle}\quad ,\quad
\chi_T=V \overline{\left\langle 
                \mathrm{tr}{\mathcal M}_T^2\right\rangle} .
\label{susceptibilities}
\end{equation}

A very useful quantity is the Binder parameter, defined as:
\begin{equation}
g^V_4=1-\frac{1}{3}\frac{\overline{\langle {\cal M}^4\rangle}}
           {\overline{\langle {\cal M}^2 \rangle}^2} \quad ,\quad
g^T_{4}=1-\frac{\overline{\langle {(\mathrm{tr}
{\mathcal M}_T^2)}^2\rangle}} 
{3 \overline{{\langle {\mathrm{tr} {\mathcal M}_T^2}\rangle}}^2} .
\label{g4cumulants}
\end{equation}

Other kind of Binder parameter, meaningless for the pure system, can be
defined as:
\begin{equation}
g^V_2=\frac{\overline{ \langle {\cal M}^2 \rangle^2 - 
\overline{\langle {\cal M}^2 \rangle}^2
}}
{\overline{ \langle {\cal M}^2 \rangle}^2 }\ ,\ 
g^T_{2}=\frac{\overline{ \langle (\mathrm{tr}
{\mathcal M}_T^2) \rangle^2 - 
\overline{\langle \mathrm{tr}{\mathcal M}_T^2 \rangle}^2
}}
{\overline{ \langle \mathrm{tr}{\mathcal M}_T^2 \rangle}^2 } ,
\label{g2cumulants}
\end{equation}
these are the quantities we are using to estimate the self-averaging properties of the susceptibility ($R_\chi$) in both channels.

A very convenient definition of the correlation length in a finite lattice 
reads, see Ref.~\cite{XIL}:
\begin{equation}
\xi=\left(\frac{\chi/F-1}{4\sin^2(\pi/L)}\right)^\frac{1}{2},
\label{XI}
\end{equation}
where $F$ is defined in terms of the Fourier transform of the
magnetization:
\begin{equation}
\boldsymbol{{\cal F}}(\boldsymbol{\mathit{k}})=\frac{1}{V}\sum_{\boldsymbol{\mathit{r}}}e^{\mathrm i
\boldsymbol{\mathit{k}\cdot\mathit{r}}} \epsilon_{\boldsymbol{\mathit r}}\boldsymbol{\mathit{S}_r}
\end{equation}
as:
\begin{equation}
F=\frac{V}{3}\overline{\left\langle |{\cal
F}(2\pi/L,0,0)|^2+\mathrm{permutations}\right\rangle}\ .
\end{equation}
The same definition is also valid in the tensorial case. This
definition is very well behaved for the finite-size scaling (FSS)
method we have employed, see Ref.~\cite{UCMOND3}. Finally, we measure
the specific heat as:
\begin{equation}
C= V^{-1} \overline{\langle {\cal E}^2 \rangle  -\langle{\cal E} \rangle ^2} \ .
\label{cesp}
\end{equation}

\section{SIMULATIONS}
\label{sec:simulations}
\subsection{Description}
\label{subsec:description}
The lattice sizes $L$ we have studied are $8,12,16,$ $24,32,48,64$,
and, only in the pure model, $L=96$.

Between each measure of the observables described in
Sec.~\ref{sec:themodel}, firstly, we update the spin-variables using a
Metropolis method over a 10\% of the individuals spins, chosen at
random, then we perform a number (growing with $L$) of cluster updates
using a Wolff method, see Ref.~\cite{amit}, this is our Elementary
Monte Carlo Step (EMCS). The number of clusters traced (or Wolff
updates) between measures have been chosen to yield a good value of
the self-correlation time, see Ref.~\cite{amit}, in our case always 
$1<\tau<2$ ($\tau$ being the integrated autocorrelation time of the energy).

In order to work in thermally-equilibrated systems 
we perform a great number of EMCS. We start the simulation
always from random (hot) distributions of the spin-variables, although
we have checked out that averages do not change if we begin from cold
configurations (i.e.~all spins being in the same direction). To
concrete, we have taken $4\times10^6$ measures for the pure model 
discarding about $10^5$ of the first measures
for $L=8$ and growing this number with the lattice size. We have
performed $2\times 10^4$ quenched disorder realizations in the diluted
models independently of the dilution and the lattice size and  taking 100
measures per sample, according to Ref.~\cite{ISDIL4D} who demonstrated
that the best approach to minimize the statistical error is to simulate
a great number of samples with a few measures in each one.

To measure the critical exponents, we use the so-called quotients
method~\cite{UCMOND3,PERC}, which allows for a great statistical
accuracy. The starting point is the equation:
\begin{equation}
\left.Q_O\right|_{Q_\xi=s}=s^{x_O/\nu}+O(L^{-\omega})\ .
\label{QUO}
\end{equation}

With $Q_O$ being the quotient of the observable $O$ measured in a pair of lattices of sizes $L$ and $sL$, at the temperature where $Q_\xi=s$, being $\omega$ the eigenvalue corresponding to the first irrelevant operator on the language of the RG theory. We have fixed $s=2$ in our case. Therefore, firstly we need to estimate by successive simulations the $\beta$ point where:
\begin{equation}
\frac{\xi(2L,\beta,p)}{2L}=\frac{\xi(L,\beta,p)}{L}\ ,
\label{BETANAIVE}
\end{equation}
for each pair of lattices $(L,2L)$, then we have used re-weighting techniques 
to fine-tune this condition. These re-weighting techniques are used to
$\beta$-extrapolate the observables and calculate their $\beta$-derivatives, always before the
sample-average is performed. The equations used are, see Refs.~\cite{amit,ISDIL4D}:
\begin{equation}
\langle {\cal O} \rangle (\beta+\Delta\beta)=
\langle {\cal O} e^{\Delta\beta{\cal E}}\rangle / \langle e^{\Delta\beta{\cal E}}\rangle\ ,
\label{FS}
\end{equation}
\begin{equation}
\partial_\beta \overline{\langle {\cal O}\rangle}=
 \overline{\partial_\beta\langle {\cal O}\rangle}=
\overline{\left\langle_{\vphantom{|}} {\cal{OE}} - \langle{\cal O}
 \rangle \langle {\cal E} \rangle\right\rangle}.
\label{DERIVADA}
\end{equation}

\begin{figure}[ht!]
\begin{center}
\includegraphics[width=0.9\columnwidth,trim=18 10 28 10]{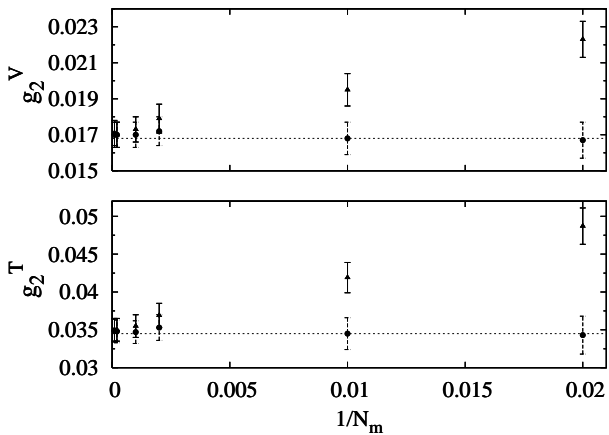}
\caption{The $g_2$ cumulant in both channels for $L=64$, $p=0.9$ with
1000 samples, $\beta_\mathrm{simulation}=0.79112$, reweighted at
$\beta=0.79082$ as a function of $1/N_m$, being $N_m$ 
the number of measures in each sample. We report data with
$N_m=50,100,500,1000,5000,10000$. The data without the bias correction
proposed in Ref.~\cite{ISDIL4D} are marked with triangles while the
corrected ones are marked with circles. We also mark with the dotted
lines the election used in this work (which corresponds to
$N_m=100$). Notice the importance of the correction of the bias if one
performs reweighting with the data.
\label{fig_extrap_Nsamples}}
\end{center}
\end{figure}

These extrapolations are biased, for instance, the expectation value of Eq.(\ref{DERIVADA}), when the averages are calculated with $N_m$ measures is:
\begin{equation}
\overline{\left(1-\frac{2\tau}{N_m}\right)\partial_\beta \langle\cal O
\rangle}\ ,
\label{BIASDER}
\end{equation}
hence, we have to correct this bias. We have followed the recipe given
in Ref.~\cite{ISDIL4D}. An example of the effect of this correction is
found in Fig.~\ref{fig_extrap_Nsamples} (for the greatest lattice size
simulated and $p=0.9$, the same applies to other dilutions and lattice
sizes): a great bias affects the non corrected numerical data and it
is clear the importance to take into account this effect. In addition,
it is clear that the recipe of Ref.~\cite{ISDIL4D} is working
perfectly for $N_m=100$, which is the number of measures per sample we
have performed in this work. Therefore, we are very confident that all
the data presented in this paper, processing with the recipe of
reference~\cite{ISDIL4D} are not biased due to the reweighting
technique.

On the other hand, we have tried to use the solution obtained in
Ref.~\cite{Hasen}, where each sample is splitted in four parts, but
the results were bad, this is due to the small number of
measures we take in each sample ($10^2$), which makes big differences
between the averages in each quarter.

To compute errors in the averages we have used Jack-Knife methods, see
Ref.~\cite{amit}. We have defined 20 Jack-Knife blocks for the pure
model in a \emph{unique} sample and one block for \emph{each} sample
in the diluted ($p<1$) models.

Calculated observables and critical exponents present
sometimes, instead of a stable value, a monotonically decreasing
one. For $\eta$, such an evolution with growing $L$ is found, but it
is clearly weaker than for $\nu$. In these cases an infinite volume
extrapolation is called for.  If hyperscaling holds, we expect
finite-volume scaling corrections proportional to $L^{-\omega}$. This
issue will be addressed in the next subsection.

\subsection{The scaling exponent $\omega$}

As can be seen in Tables from~\ref{expmagpuro}~to~\ref{expter05},
allocated in the appendix A, specially for the thermal exponents and
the cumulants ($g_4$ and $g_2$), there are evident finite volume
effects so we have to use the equation:
\begin{equation}
\left.\frac{x_O}{\nu}\right|_{\infty} -\left.\frac{x_O}{\nu}
\right|_{(L,2L)}\propto L^{-\omega},
\label{Xw}
\end{equation}
which is a consequence of scale-hypothesis, first derived in
Ref.~\cite{HIPERSCALA}. Consequently, choosing a good value for
$\omega$ is a primordial question.

Exact results and RG calculations tell us that the disorder, being
irrelevant in this model, induce scaling corrections with an exponent
$\alpha/\nu\simeq -0.188$ (in $L$)~\cite{PeliVica}. In addition to
this new scaling correction one must have the one of the pure model,
which is related to the coupling of the $(\phi^2)^2)$ term in the
Ginzburg-Landau theory. This exponent is assumed to be
$0.8$~\cite{guida,hasen2} (for the pure model). Hence, the leading one
is the exponent induced by the disorder. We will try to check this
scenario by computing the ``leading'' correction to the scaling
exponent from the numerical data.

First of all, we have tried to estimate $\omega$ just by considering
it as another tunable parameter in Eq.~(\ref{Xw}) applied to some
physical quantities. In these fits, as first approximation, we
disregarded the possible correlations between the data for different
$L$ values, the results can be seen in
Table~\ref{omega_weightmean}. If we perform a weighted average with
this results we obtain $\omega=1.07(9)$ for the pure model and
$\omega=0.92(9)$, $\omega=0.81(7)$ and $\omega=0.88(4)$ for the
diluted model with $p=0.9,0.7$ and $0.5$ respectively, in very good
agreement with the value from the references given before. However, we
think this method is not very confident because of the variability of
the results from a quantity to another as shown in the
Table~\ref{omega_weightmean}.
\begin{table}[ht]
\begin{center}
\begin{tabular}{|c|c|c|c|c|}\hline
$\cal O$ & $\omega_{p=1.0}$ & $\omega_{p=0.9}$ & $\omega_{p=0.7}$& $\omega_{p=0.5}$\\\hline\hline
$\eta_{\chi^V}$               & 1.45(52)  & ---      & ---       & ---\\\cline{1-5}
$\eta_{M^V}$                  & 1.62(80)  & ---      & ---       & ---\\\cline{1-5}
$\eta_{\chi^T}$               & ---       & ---      & 1.2 (1.1) & 0.68(46) \\\cline{1-5}
$\eta_{M^T}$                  & ---       & ---      & ---       & 0.73(46) \\\cline{1-5}
$\nu_{\partial_\beta g_4^V}$  & ---       & ---      & ---       & ---\\\cline{1-5}
$\nu_{\partial_\beta\xi^V}$   & 2.30(61)  & ---      & ---       & 0.62(47)\\\cline{1-5}
$\nu_{\partial_\beta g_4^T}$  & ---       & ---      & ---       & --- \\\cline{1-5}
$\nu_{\partial_\beta\xi^T}$   & 2.12(52)  & 1.76(60) & 1.09(40)  & 1.34 (27)\\\cline{1-5}
$\xi^V/L$                     & 1.08(21)  & 1.21(31) & 0.61(12)  & 0.45(10)\\\cline{1-5}
$\xi^T/L$                     & ---       & ---      & 1.55(76)  & 1.64(17)\\\cline{1-5}
$g_4^V$                       & 0.85(14)  & 2.00(61) & 1.21(15)  & 1.19(13)\\\cline{1-5}
$g_4^T$                       & 1.06(14)  & ---      & 1.35(33)  & 1.41(42)\\\cline{1-5}
$g_2^V$                       & ---       & 0.81(16) & 0.89(9)   & 0.94(7)\\\cline{1-5}
$g_2^T$                       & ---       & ---      & 0.63(12)  & 0.72(10)\\\hline\hline
$\bar{\omega}_\mathrm{weighted}$     & 1.07(9)   & 0.92(9) & 0.81(7)   & 0.88(4)\\\hline
\end{tabular}
\caption{$\omega$ values from the $L\rightarrow\infty$
extrapolations of some quantities. In the last row can be seen the
weighted average of each column. We disregarded data with error bars
bigger than the 100\% of the values themselves. Those disregarded data
are showed in the table as ---.}
\label{omega_weightmean}
\end{center}
\end{table}

Another approach, following Ref.~\cite{ISDIL3D}, is to study the
crossing points of scaling functions (as $\xi/L$ and $g_4$) measured
in pairs of lattices with sizes $L$ and $2L$. The deviation of these
crossing point from the infinite volume critical coupling will behave
as:
\begin{equation} 
\Delta\beta(L,sL)\equiv \\
\beta(L,sL)-\beta_{\mathrm{c}}(\infty) \propto
\frac{1-s^{-\omega}}{s^{\frac{1}{\nu}}-1}L^{-\omega-\frac{1}{\nu}}.
\label{BETADESVIATION}
\end{equation} 
With this method we need an additional estimate for the thermal
exponent $\nu$, we have used,
following~\cite{peli}, the value
$\nu=0.7113(11)$ for the pure model (notice the really small error in
$\nu$, so we will discard it in the following, see the comment at the
end of this section), which is also a valid value for the diluted
models, because of the validity of the Harris Criterion and as can be
checked with the data in the Appendix~A. Again we fixed $s=2$. In this
approach, we only use the crossing points in the vectorial channel
because they are cleaner.

\begin{figure}[ht]
\begin{center}
\includegraphics[width=\columnwidth,trim=18 10 18 10]{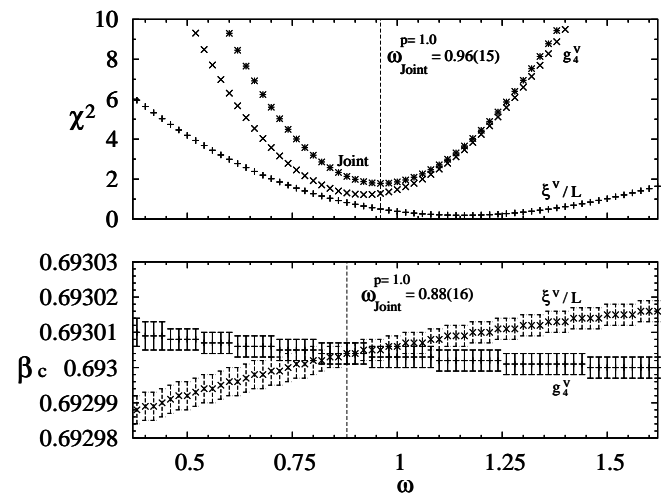}
\caption{ \emph{Upper part:} $\chi^2$ as a function of $\omega$
deduced from the fits to $L\rightarrow\infty$, Eq.~(\ref{BETADESVIATION}),
for the crossing point of $\xi/L$ and $g_4$ for the ($L$,\ $2L$) pair for
the pure model. Also is shown the joined $\chi^2$. \emph{Lower part:}
extrapolated $\beta_c(\infty)$ as a $\omega$ function, the point where
both observables give the same extrapolated value is marked with the
dotted line.
\label{calculoomega_puro}}
\end{center}
\end{figure}

\begin{figure}[ht]
\begin{center}
\includegraphics[width=0.9\columnwidth,trim=18 10 28 10]{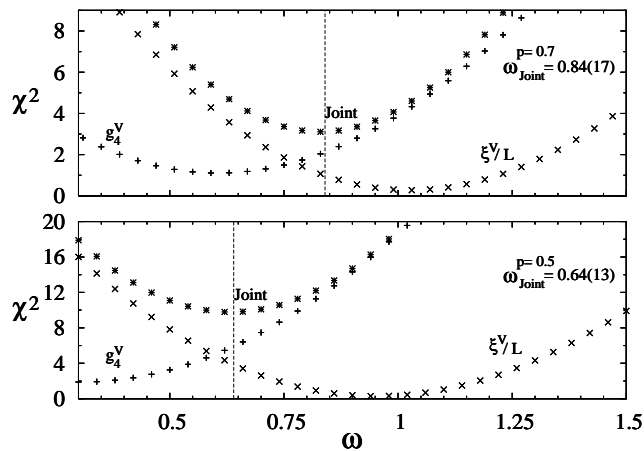}
\caption{ \emph{Upper part:} $\chi^2$ as a function of $\omega$ for
the diluted ($p=0.7$) model. Also is shown the joined
$\chi^2$. \emph{Lower part:} $\chi^2$ as a function of $\omega$ for
the diluted ($p=0.5$) model.
\label{calculoomega_diluidos}}
\end{center}
\end{figure}

Extrapolating this crossing points using Eq.(\ref{BETADESVIATION}), we
can plot the minimum of the $\chi^2$ of the fit as a function of
$\omega$ obtaining the upper part of Fig.~\ref{calculoomega_puro}~and
the whole Fig~\ref{calculoomega_diluidos}. To carry out this
extrapolations we have to realize that the measures of the crossing
points are correlated by pairs, so we have to use the $\chi^2$
definition that include the whole self-covariance matrix:
\begin{equation} 
\chi_x^2\ =\ \sum_{l=1}^{N} \sum_{m=1}^{N} (x_l-\mathrm{fit})(\mathrm{cov}^{-1})_{l,m}(x_m-\mathrm{fit}) ,
\label{CHISQU} 
\end{equation} 
being $N$ the number of crossing points, it is to say, the number of
simulated $L$ values minus two, $x_{l}$ is the obtained value for the
observable $x$ ($\xi^V/L$ or $g^V_4$), at the crossing point for $L_l$
and $2L_l$, and ``$\mathrm{fit}$'' is the fitted value to the form of
Eq.~(\ref{BETADESVIATION}) for $L_l$. In addition:
\begin{equation} 
\mathrm{(cov)}_{l,m}=\langle x_m x_l \rangle - \langle x_m \rangle
\langle x_l \rangle
\label{COV} 
\end{equation} 
can be also defined in terms of Jack-Knife blocks, see Ref.~\cite{amit}, as:
\begin{equation} 
\mathrm{(cov)}_{l,m}=\frac{N_{b}-1}{N_{b}} \sum_{i=1}^{N_{b}}
(x_{l,i}^{\mathrm{J-K}}- \langle x_l \rangle)(x_{m,i}^{\mathrm{J-K}}
- \langle x_m \rangle)  .
\label{COV_JK} 
\end{equation} 
where $N_b$ is the number of Jack-Knife blocks,
$x_{l,i}^{\mathrm{J-K}}$ are block variables, where the first subindex
runs over $L$ values while the second one over J-K blocks, and
$\langle x_l \rangle$ is the average between all block variables given
$L=L_l$.

Also, following Ref.~\cite{ISDIL3D}, we can do a joined
fit in $\omega$ of the crossing points of $\xi^V/L$ and $g^V_4$ by defining:
\begin{equation} 
\chi^2_\mathrm{joined}=\chi^2_{\xi^V/L}+\chi^2_{g^{V}_4}
\label{CHICUAD_JOINT} 
\end{equation} 
using Eq.~(\ref{CHISQU}) to calculate each one of the latter terms and
searching for the minimum of $\chi^2_\mathrm{joined}$. We can obtain
the error in $\omega$ searching for the point
$\omega_1$ in which
$\chi^2_\mathrm{joined}(\omega_1)=
\chi^2_\mathrm{joined}(\omega_\mathrm{min})+1$,
so the error  is
$\Delta\omega=|\omega_\mathrm{min}-\omega_1|$. The results for this
joined fits are shown in the upper part of
Fig.~\ref{calculoomega_puro} and in the whole
Fig.~\ref{calculoomega_diluidos}. We find with this method the values:
\begin{center}
$\omega=0.96(15) ,$ 
\end{center}
for the pure model and:
\begin{center}
 $\omega=2.29(70)\ ,\ \ 0.84(17)\ ,\ \ 0.64(13)$, 
\end{center}
for the diluted models with $p=0.9,\ 0.7\ \mathrm{and}\ 0.5$
respectively, in agreement with the value obtained in the pure model
~\cite{peli,UCMOND3,guida,hasen2}, except in the $p=0.9$ case, in
which the value is far away two standard deviations from
$\omega=0.8$~\cite{guida,hasen2}.  One possibility is that we are
computing the leading correction to scaling exponent but with a large
error. Another possibility is that in the $p=0.9$ model the
coefficients of the leading correction (from the numerical point of
view) to the scaling vanishes or are very small.  This result and the
change in the slope of the $g_4$ data for $p<1$ respect to the $p=1$
ones, as can be seen in Table~\ref{cumulantes_todos}, are evidences of
the possible \emph{improved action} found for $p=0.9$, see
Ref.~\cite{Hasen}, therefore the $\omega$ exponent we are measuring in
this case could correspond to the third irrelevant operator, instead
of the second one (remember that following RG the first one is
$\alpha/\nu\simeq -0.188$).

In addition, as Ref.~\cite{ISDIL3D} also did, we could estimate the
correct value for $\omega$ as the one producing the same
$\beta_c(\infty)$ value for $\xi/L$ and $g_4$, as can be seen in
the lower part of Fig.~\ref{calculoomega_puro}, marked with the dotted
line in $\omega=0.88$. This approach only works for the pure model in
which such a point is found, with another $p$ value the
$\beta_c(\infty)$ estimates from $\xi/L$ and $g_4$ do not cross each
other.

In conclusion, we have shown that our data (both for the pure and
diluted model) are fully compatible with the value $\omega=0.80(1)$,
obtained previously both numerically and analytically for the pure
model~\footnote{In more detail, the field theoretical approaches (both
fixed dimension and $\epsilon$-expansion) provide very accurate values
for $\omega$: 0.782(13) and 0.794(18)
(respectively)~\cite{guida}. Recent numerical simulations bring us the
values 0.775(13) and 0.799(13)~\cite{hasen2} and 0.64(13) and
0.71(15)~\cite{UCMOND3}.}. In addition, since the error bars in
$\omega$ are really small (one per cent of error) we have discarded
the uncertainty in $\omega$ in the analysis presented in this paper,
the error bars in the extrapolated quantities are much bigger than the
uncertainty caused by the error bars in $\omega$: so we have fixed
$\omega=0.80$.  The extrapolations (Tables III to XIII) and Figures (3
to 8) shown in the rest of the paper are obtained using this value of
$\omega$.

Finally, it is interesting to note that we have seen in the
analysis presented in this subsection no traces of the leading
correction to the  scaling exponent even for the strongest dilution we
have simulated, which should be $\alpha/\nu \simeq -0.188$. One can
explain this fact assuming that the amplitudes of this scaling
correction exponent are really small, so we are seeing the next to the
leading scaling correction.

\subsection{Numerical Results about Self-Averaging}
\label{subsec:numresults}

Once that we have checked that the value $\omega=0.80$ describe the
corrections to the scaling in the pure and diluted model we can try to
extrapolate the values of the two $g_2$ to infinite volume.

Numerical results for $g_2$ and $g_4$ in both channels are shown in
Table~\ref{cumulantes_todos} both for pure (only $g_4$) and diluted
models.
\begin{table}[ht]
\begin{center}
\begin{tabular}{|c|c|c|c|c|c|}\cline{1-6}
$p$& $L$ & $g_2^V$ & $g_2^T$ & $g_4^V$ & $g_4^T$ \\\hline\hline
1.0 & 8 & 0 & 0 & 0.62243(4) & 0.5216(1) \\
   & 12 & 0 & 0 & 0.62172(5) & 0.5189(2) \\
   & 16 & 0 & 0 & 0.62152(6) & 0.5181(2) \\
   & 24 & 0 & 0 & 0.62100(5) & 0.5166(2) \\
   & 32 & 0 & 0 & 0.62092(3) & 0.5162(1) \\
   & 48 & 0 & 0 & 0.62066(5) & 0.5156(2) \\\cline{1-6}

0.9 & 8 & 0.0327(4) & 0.0576(7)  & 0.6151(2) & 0.5102(3)\\
   & 12 & 0.0273(3) & 0.0518(6)  & 0.6163(1) & 0.5104(3) \\
   & 16 & 0.0253(3) & 0.0499(6)  & 0.6166(1) & 0.5100(3) \\
   & 24 & 0.0226(3) & 0.0453(6)  & 0.6168(1) & 0.5098(3) \\
   & 32 & 0.0208(2) & 0.0421(5)  & 0.6171(1) & 0.5100(3) \\\cline{1-6}

0.7 & 8 & 0.0780(8) & 0.1406(16)  & 0.6061(3) & 0.4994(6)\\
   & 12 & 0.0610(6) & 0.1177(13)  & 0.6108(2) & 0.5039(5) \\
   & 16 & 0.0512(5) & 0.1009(11)  & 0.6131(2) & 0.5064(4) \\
   & 24 & 0.0423(4) & 0.0868(10)  & 0.6150(2) & 0.5077(4) \\
   & 32 & 0.0371(4) & 0.0770(9)   & 0.6160(2) & 0.5089(4) \\\cline{1-6}

0.5 & 8 & 0.1130(11) & 0.2061(24) & 0.6006(4) & 0.4999(8)\\
   & 12 & 0.0834(8)  & 0.1600(18) & 0.6072(3) & 0.5047(6) \\
   & 16 & 0.0702(7)  & 0.1395(16) & 0.6107(3) & 0.5070(6) \\
   & 24 & 0.0553(6)  & 0.1138(13) & 0.6138(2) & 0.5085(5) \\
   & 32 & 0.0474(5)  & 0.0980(11) & 0.6151(2) & 0.5095(4) \\\cline{1-6}

\end{tabular}
\caption{Cumulants for the O($3$) model. In the first column is
represented the spin density $p$. All the cumulants are calculated in
the crossing points of $\xi/L$ for $L$ and $2L$. The averages have
been computing using $10^4$ samples (except in the $p=1$) case.}
\label{cumulantes_todos}
\end{center}
\end{table}

First of all, we will try to check the non zero $g_2$ scenario with
the correction to the scaling exponent fixed to that obtained in the
previous section.

\begin{figure}[htbp]
\begin{center}
\includegraphics[width=\columnwidth,trim=18 10 18 10]{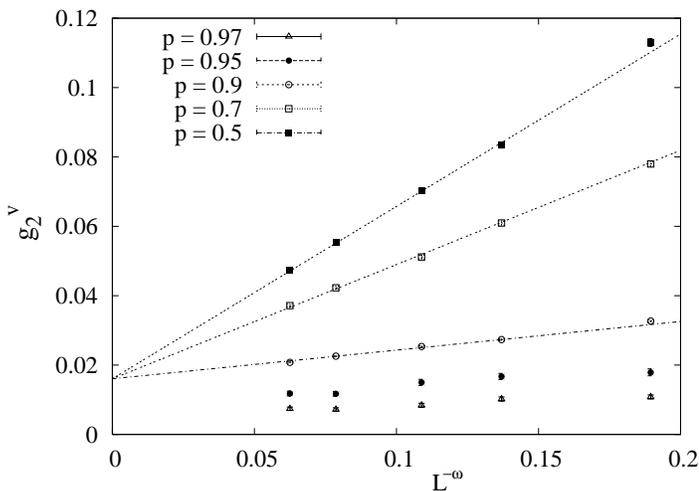}
\caption{ Joined extrapolation to $L\rightarrow\infty$ for the $g_2$
cumulant of the vectorial susceptibility. Extrapolations are carried
out by choosing a common value for the first term of Eq.~(\ref{Xw})
for all dilutions and by minimizing the joined $\chi{^2}$. We have
disregarded the data with $L=8$ to obtain a good value of the
$\chi^2$.}
\label{fig:extrap_g2V}
\end{center}
\end{figure}

\begin{figure}[htbp]
\begin{center}
\includegraphics[width=\columnwidth,trim=18 10 18 10]{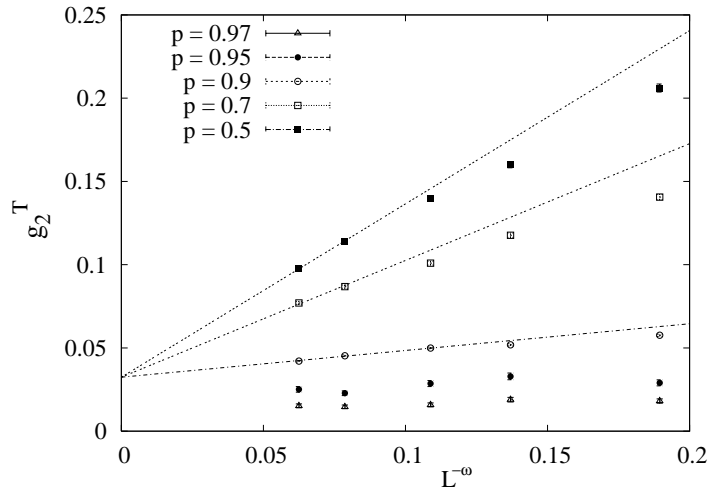}
\caption{ Joined extrapolation to $L\rightarrow\infty$ for the $g_2$
cumulant of the tensorial susceptibility, to the form of Eq.~(\ref{Xw}).}
\label{fig:extrap_g2T}
\end{center}
\end{figure}

We have found that it is possible to extrapolate using the form of
Eq.~(\ref{Xw}) (performing a joint fit~\footnote{We have used to do
all fits the data set which gives out the smallest $\chi^2$ with the
biggest number of degrees of freedom.}) the values of $g_2$ to a value
(depending only on the channel), which is independent to the dilution,
and near the analytical prediction of reference \cite{korut}. However,
simulations (with a small number of samples), at dilutions $p=0.95$
and $p=0.97$ do not follow the scaling found for $p\le 0.90$ (see
Table~\ref{cumulantes_pgrande}). Hence, our numerical data do not
support the scenario $g_2\neq 0$, see Figs.~\ref{fig:extrap_g2V}
and~\ref{fig:extrap_g2T}. Notice, see also
Table~\ref{cumulantes_pgrande}, that all the values for these two
lowest dilutions are smaller than the extrapolated point and they are
decreasing (for both channels and taking into account the error bars).

\begin{figure}[htbp]
\begin{center}
\includegraphics[width=\columnwidth,trim=18 10 18 10]{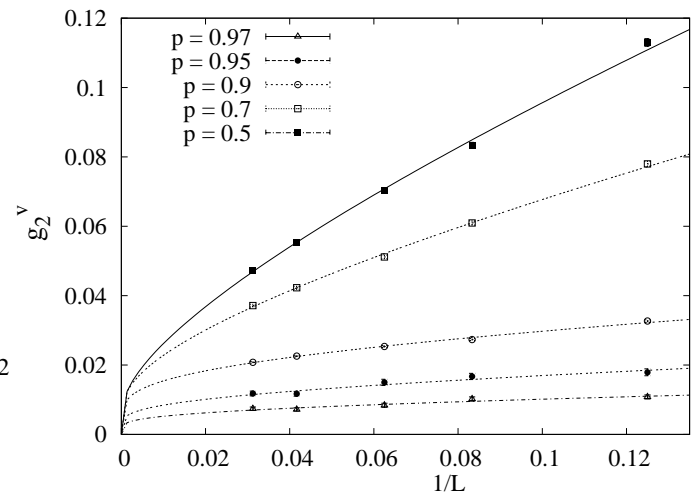}
\caption{Extrapolation to $L\rightarrow\infty$ for the $g_2$
cumulant of the vectorial susceptibility. The fitting function is in this case
of the form $g_2 = aL^{\alpha/\nu}+bL^{-\omega}$.}
\label{fig:extrap_g2V_pelissetto}
\end{center}
\end{figure}

\begin{figure}[htbp]
\begin{center}
\includegraphics[width=\columnwidth,trim=18 10 18 10]{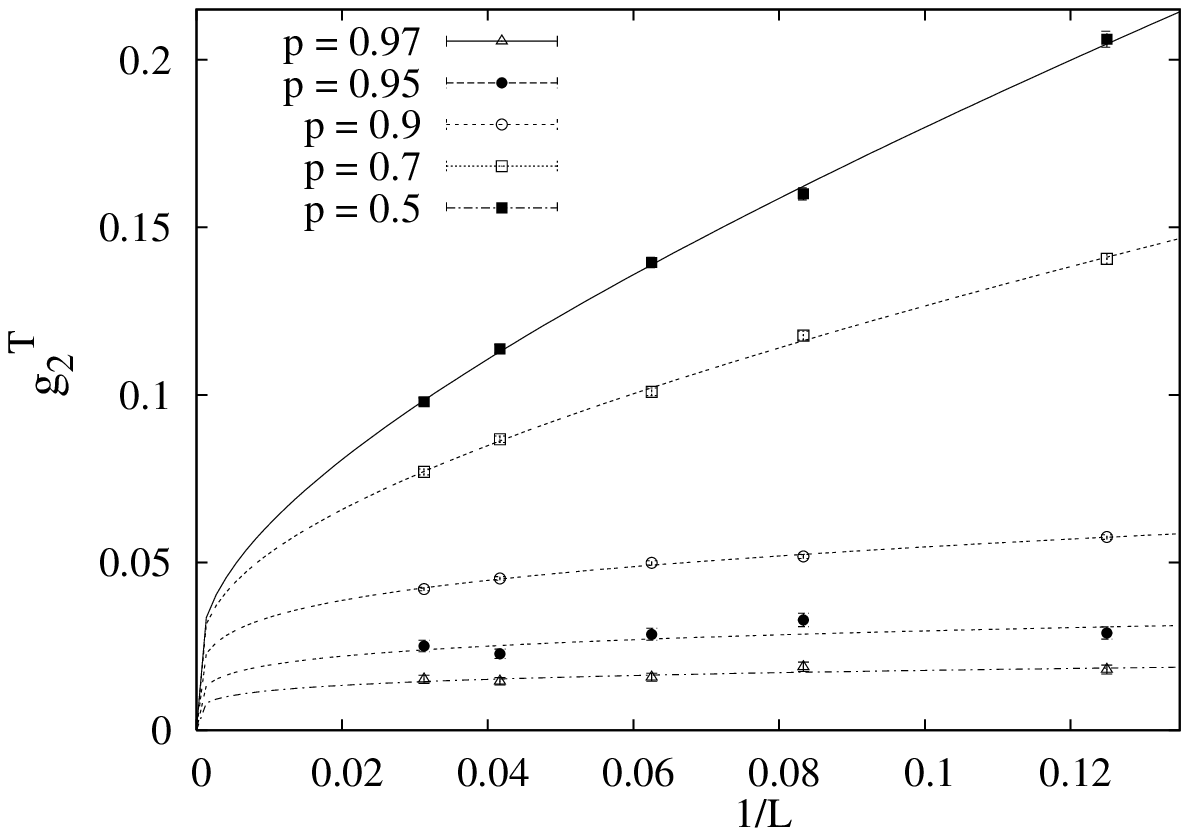}
\caption{Extrapolation to $L\rightarrow\infty$ for the $g_2$
cumulant of the tensorial susceptibility to the form $g_2 = aL^{\alpha/\nu}+bL^{-\omega}$.}
\label{fig:extrap_g2T_pelissetto}
\end{center}
\end{figure}

Secondly, we will check the $g_2=0$ scenario. To do this, we
extrapolate $g_2$ using the form proposed in Ref.~\cite{harris} ($g_2
\backsim L^{\alpha/\nu}$) {\em but} also including the term
$L^{-\omega}$:
\begin{equation}
g_2=a L^{\alpha/\nu} + b L^{-\omega}\,.
\end{equation}
We obtain the fits shown in Figs.~\ref{fig:extrap_g2V_pelissetto}
and~\ref{fig:extrap_g2T_pelissetto} for both channels.  The $\chi^2$ of
these all fits are really good, hence, we have obtain strong evidence
supporting this $g_2=0$ scenario. Notice that the introduction of the
scaling correction has had paramount importance in order to obtain
very good $\chi^2$ in all the fits. The numerical data, for the lattice
size simulated, does not follow the one term dependence $g_2 \propto L
^{\alpha/\nu}$.

\begin{table}[ht]
\begin{center}
\begin{tabular}{|c|c|c|c|c|c|}\cline{1-6}
$p$& $L$ & $g_2^V$ & $g_2^T$ & $g_4^V$ & $g_4^T$ \\\hline\hline
0.97 & 8  & 0.0108(6) & 0.0181(13)  & 0.6201(4) & 0.5187(10) \\
     & 12 & 0.0102(6) & 0.0189(14)  & 0.6195(4) & 0.5164(10) \\
     & 16 & 0.0084(6) & 0.0158(12)  & 0.6201(4) & 0.5159(10) \\
     & 24 & 0.0072(5) & 0.0146(11)  & 0.6206(4) & 0.5162(9)  \\
     & 32 & 0.0074(5) & 0.0152(12)  & 0.6206(4) & 0.5152(10) \\\cline{1-6}

0.95 & 8  & 0.0179(10) & 0.0290(18) & 0.6180(5) & 0.5158(11)\\
     & 12 & 0.0167(9)  & 0.0329(20) & 0.6182(5) & 0.5116(12) \\
     & 16 & 0.0150(9)  & 0.0286(18) & 0.6181(5) & 0.5129(11) \\
     & 24 & 0.0117(7)  & 0.0228(14  & 0.6186(4) & 0.5135(11) \\
     & 32 & 0.0118(7)  & 0.0251(17) & 0.6193(4) & 0.5140(10) \\\cline{1-6}

\end{tabular}
\caption{Cumulants for the O($3$) model with high $p$ values. All the cumulants are computing averaging only  $10^3$ samples.}
\label{cumulantes_pgrande}
\end{center}
\end{table}

\section{CONCLUSIONS}
\label{sec:conclusions}
We have studied the critical properties of the Heisenberg diluted model
for different values of the dilution using the quotient method. Our
main aim was to check the self-averaging properties of the
susceptibility.

We have studied in a great detail the corrections to the scaling in
the diluted Heisenberg model. We have obtained that the numerical data
follow the next to the leading correction to the scaling exponent
instead the leading one. We will show in the appendix all the critical
exponents and cumulants using this next to the leading exponent, also
we report that the result of this analysis is fully 
compatible with the RG predictions and the Harris Criterion: our
exponents and cumulants are compatibles with that of the pure model
and independent of the dilution with a high degree of precision.

In addition, we have shown that we obtain non universal quantities if
we assume $\alpha/\nu$ as the main scaling correction even if we add
the $\omega$ correction to the scaling exponent (see the appendix),
using two correction to scaling exponents in the analysis.

Finally, we have shown strong evidence for a zero $g_2$ cumulant, both
in vector and tensor channels, in the thermodynamic limit at
critically, contrasting with some analytical predictions~\cite{korut},
and in agreement with the ones obtained in Ref.~\cite{harris}.  The
introduction of scaling corrections in the analysis has become crucial
to obtain the $g_2=0$ scenario. In addition, simulations of samples
with lower dilution have helped to discard the $g_2 \neq 0$ scenario.

\begin{acknowledgments}
This research has been supported by the Ministerio de Educaci\'on y
Ciencia (Spain) through grant No.\ FIS2004-01399 (partially financed
by FEDER funds), through grant No.\ BFM2003-C08532 and by the European
Community's Human Potential Programme under contract
HPRN-CT-2002-00307. The computations have been carried out using the
resources of the BIFI (Instituto de Biocomputaci\'on y F\'{\i}sica de
Sistemas Complejos) placed in Zaragoza, Spain.  We thank A. Pelissetto
and E. Vicari for pointing us the relevance of the scaling corrections
in the analysis of the $g_2=0$ scenario. We also have to thank
L.~A.~Fernandez, V.~Mart\'{i}n~Mayor, and E.~Korutcheva for
interesting discussions.
\end{acknowledgments}

\appendix
\section*{Appendix A: Critical Exponents and Cumulants}

In this appendix we will check the consistency of the $\omega$
exponent computed in the text by means of the computation of critical
exponents and cumulants. In addition, we will check if these sets of
exponents are Universal or not by comparing different dilutions
with the pure model. We will use in this analysis the data from
$p=0.9, 0.7$ and 0.5 (we have simulated in these values of the dilution $10^4$
samples).

\begin{figure}[!ht]
\begin{center}
\includegraphics[width=\columnwidth,trim=18 10 18 10]{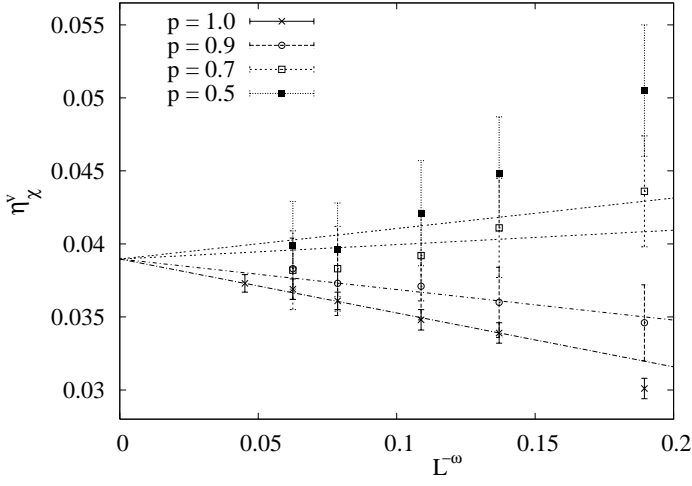}
\caption{Joined extrapolation to $L\rightarrow\infty$ for the $\eta$ exponent deduced from the vectorial susceptibility ($\chi^V$). Extrapolations are carried out by choosing a common value for the first term of Eq.~(\ref{Xw}) for all dilutions, and by minimizing the joined $\chi^2$.}
\label{fig:extrap_eta_susV}
\end{center}
\end{figure}

\begin{figure}[!ht]
\begin{center}
\includegraphics[width=\columnwidth,trim=18 10 18 10]{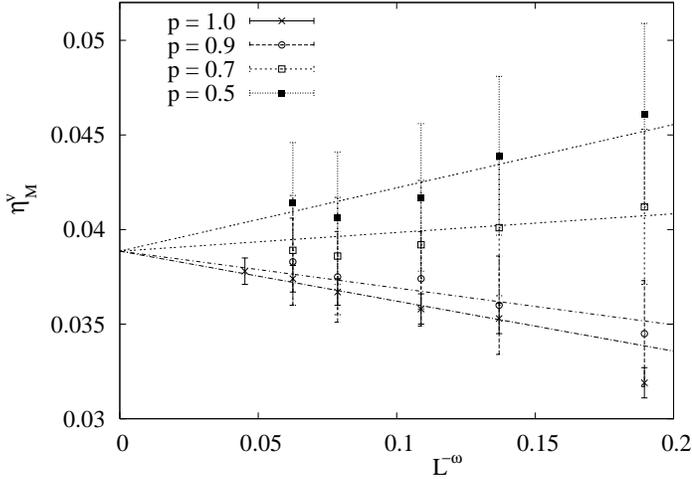}
\caption{ Joined extrapolation with all $p$ values to $L\rightarrow\infty$ for the $\eta$ exponent deduced from  the vectorial magnetization ($M^V$). }
\label{fig:extrap_eta_magV}
\end{center}
\end{figure}

\begin{figure}[!ht]
\begin{center}
\includegraphics[width=\columnwidth,trim=18 10 18 10]{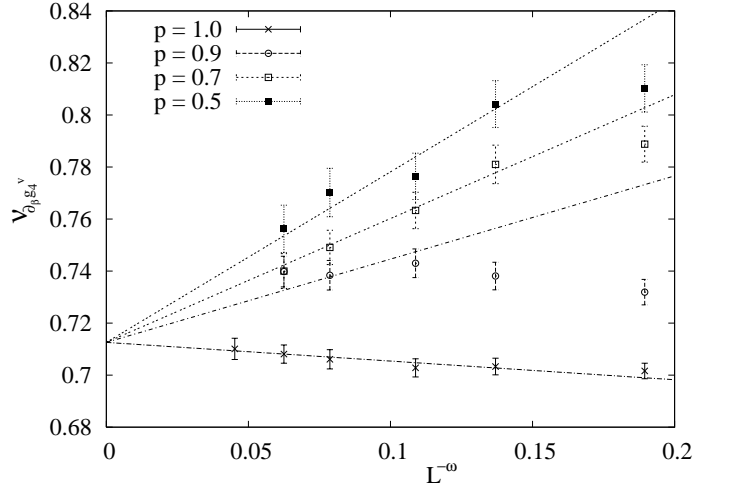}
\caption{ Joined extrapolation with all $p$ values to $L\rightarrow\infty$ for the $\nu$ exponent deduced from the $\partial_\beta g_4^V$.}
\label{fig:extrap_nu_dg4V}
\end{center}
\end{figure}

\begin{figure}[!ht]
\begin{center}
\includegraphics[width=\columnwidth,trim=18 10 18 10]{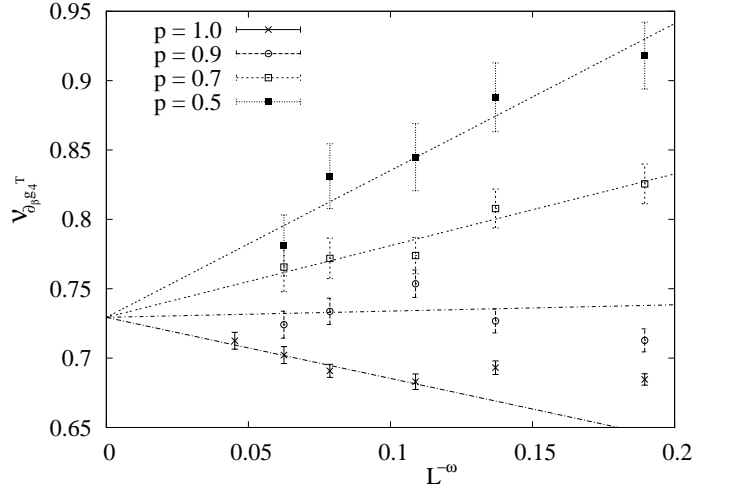}
\caption{Joined extrapolation to $L\rightarrow\infty$ for the $\nu$ exponent deduced from the $\partial_\beta g_4^T$.}
\label{fig:extrap_nu_dg4T}
\end{center}
\end{figure}

Eq.~(\ref{QUO}) applied to the operators $\partial_\beta \xi$,
$\partial_\beta g_4$, $M$ and $\chi$, yields respectively the critical
exponents $1+1/\nu$, $1/\nu$, $(d-2+\eta)/2$ and $2-\eta$.  The
numerical results are shown in Tables~\ref{expmagpuro}
and~\ref{expterpuro} for the pure model, Tables~\ref{expmag09}
and~\ref{expter09} for $p$~=~0.9, Tables~\ref{expmag07}
and~\ref{expter07} for~$p$~=~0.7 and Tables~\ref{expmag05}
and~\ref{expter05} for~$p$~=~0.5.  We have also carried out a joined
extrapolation for every $p$ values by fixing the same value of the
extrapolated exponent (first term in Eq.~(\ref{Xw})) for every $p$ value
and then minimizing the joined $\chi^2$. Some of these fits are shown in
figures~\ref{fig:extrap_eta_susV}~to~\ref{fig:extrap_nu_dg4T} and
the compared results can be seen in Tables~\ref{comparationmag}
and~\ref{comparationtherm}.

The joined extrapolation of the Binder cumulant $g_4$ is shown in
Table~\ref{comparationbind} and the agreement of our
results with the ones obtained in Refs.~\cite{peli} (numerical for the
pure model) and \cite{korut} (analytical) is really very good.
We obtain also complete agreement with previous numerical estimates of the pure
model critical exponents, see Ref.~\cite{peli}.

We obtain non universal critical exponents and cumulants if instead
$\omega=0.8$ we use $\omega=-\alpha/\nu$ as the correction to scaling
exponent. In addition, the dilution dependent exponents and cumulants
are clearly different of the pure ones. Furthermore, this scenario does
not change if we fit the data using both $\omega=-\alpha/\nu$ and
$\omega=0.8$.

\begin{table}[!ht]
\begin{center}
\begin{tabular}{c|c|c|c|c|}\cline{2-5}
& \multicolumn{2}{c|}{$\eta$} & \multicolumn{2}{c|}{$\eta_T$}\\
\cline{1-5}
\multicolumn{1}{|c|}{$L$}& \multicolumn{1}{c|}{$\chi$}      
    & \multicolumn{1}{c|}{\lower2pt\hbox{$M$}}
    & \multicolumn{1}{c|}{$\chi_T$}        
    & \multicolumn{1}{c|}{\lower2pt\hbox{$M_T$}}\\\hline\hline

\multicolumn{1}{|c|}{8}  & 0.0301(7) & 0.0319(8) & 1.4301(12) & 1.4343(13)\\\cline{1-5}
\multicolumn{1}{|c|}{12} & 0.0339(7) & 0.0353(8) & 1.4324(11) & 1.4352(12)\\\cline{1-5}
\multicolumn{1}{|c|}{16} & 0.0348(7) & 0.0358(8) & 1.4310(11) & 1.4335(12)\\\cline{1-5}
\multicolumn{1}{|c|}{24} & 0.0361(6) & 0.0367(7) & 1.4293(9)  & 1.4307(10)\\\cline{1-5}
\multicolumn{1}{|c|}{32} & 0.0369(7) & 0.0374(7) & 1.4289(11) & 1.4300(12)\\\cline{1-5}
\multicolumn{1}{|c|}{48} & 0.0373(6) & 0.0378(7) & 1.4271(9)  & 1.4280(10)\\\hline\hline
\multicolumn{1}{|c|}{$L\rightarrow\infty$}   & 0.0391(9)& 0.0390(10)& 1.4250(13)& 1.4249(15)\\\cline{1-5}
\multicolumn{1}{|c|}{$\mathrm{\chi^2/d.o.f}$}& 0.138/3  & 0.354/3   & 1.047/3   & 1.952/3   \\\cline{1-5}
\multicolumn{1}{|c|}{$\mathrm{prob}$}        & 0.987    & 0.950     & 0.790     & 0.582     \\\hline

\end{tabular}
\caption{Magnetic exponents for the pure O($3$) model. The last three rows correspond to the $L\rightarrow\infty$ extrapolation (disregarding data with $L=8$)}
\label{expmagpuro}
\end{center}
\end{table}

\begin{table}[!ht]
\begin{center}
\begin{tabular}{c|c|c|c|c|}\cline{2-5}
& \multicolumn{4}{c|}{$\nu$} \\
\cline{1-5}
\multicolumn{1}{|c|}{$L$} & \multicolumn{1}{c|}{$\partial_\beta g_4^V$}      
    & \multicolumn{1}{c|}{$\partial_\beta \xi^V$} 
    & \multicolumn{1}{c|}{$\partial_\beta g_4^T$}        
    & \multicolumn{1}{c|}{$\partial_\beta \xi^T$}\\\hline\hline
        
\multicolumn{1}{|c|}{8}  & 0.7016(30) & 0.7217(13) & 0.6846(41) & 0.7306(14)\\\cline{1-5} 
\multicolumn{1}{|c|}{12} & 0.7033(32) & 0.7162(14) & 0.6931(49) & 0.7188(13)\\\cline{1-5} 
\multicolumn{1}{|c|}{16} & 0.7028(35) & 0.7123(16) & 0.6830(56) & 0.7118(17)\\\cline{1-5} 
\multicolumn{1}{|c|}{24} & 0.7061(37) & 0.7123(17) & 0.6908(47) & 0.7112(18)\\\cline{1-5} 
\multicolumn{1}{|c|}{32} & 0.7081(35) & 0.7121(19) & 0.7022(61) & 0.7116(23)\\\cline{1-5} 
\multicolumn{1}{|c|}{48} & 0.7101(41) & 0.7118(19) & 0.7125(61) & 0.7085(21)\\\hline\hline

\multicolumn{1}{|c|}{$L\rightarrow\infty$}    & 0.7109(38)& 0.7071(19)& 0.7082(51) & 0.7071(35)\\\cline{1-5}
\multicolumn{1}{|c|}{$\mathrm{\chi^2/d.o.f}$} & 0.667/4   & 4.104/4   & 7.039/4    & 0.565/2 \\\cline{1-5}
\multicolumn{1}{|c|}{$\mathrm{prob}$}         & 0.954     & 0.392     & 0.134      & 0.754 \\\hline  

\end{tabular}
\caption{Thermal critical exponents for the pure O($3$) model. In the last column we have disregarded data with $L<16$.}
\label{expterpuro}
\end{center}
\end{table}

\begin{table}[!ht]
\begin{center}
\begin{tabular}{c|c|c|c|c|}\cline{2-5}
& \multicolumn{2}{c|}{$\eta$} & \multicolumn{2}{c|}{$\eta_T$}\\\cline{1-5}
\multicolumn{1}{|c|}{$L$}& \multicolumn{1}{c|}{$\chi$}
    & \multicolumn{1}{c|}{\lower2pt\hbox{$M$}}
    & \multicolumn{1}{c|}{$\chi_T$}
    & \multicolumn{1}{c|}{\lower2pt\hbox{$M_T$}}\\\hline\hline

\multicolumn{1}{|c|}{8}  & 0.0346(26) & 0.0345(28) & 1.4154(36) & 1.4176(37) \\\cline{1-5}
\multicolumn{1}{|c|}{12} & 0.0360(24) & 0.0360(26) & 1.4195(34) & 1.4207(36) \\\cline{1-5}
\multicolumn{1}{|c|}{16} & 0.0371(23) & 0.0374(25) & 1.4207(34) & 1.4218(35) \\\cline{1-5}
\multicolumn{1}{|c|}{24} & 0.0373(22) & 0.0375(24) & 1.4204(32) & 1.4221(34) \\\cline{1-5}
\multicolumn{1}{|c|}{32} & 0.0383(21) & 0.0383(23) & 1.4219(31) & 1.4227(33) \\\hline\hline
                                                
\multicolumn{1}{|c|}{$L\rightarrow\infty$}   & 0.0397(29) & 0.0399(31) & 1.4245(41) & 1.4252(43)\\\cline{1-5}
\multicolumn{1}{|c|}{$\mathrm{\chi^2/d.o.f}$}& 0.292/3    & 0.124/3    & 0.544/3    & 0.137/3 \\\cline{1-5}
\multicolumn{1}{|c|}{$\mathrm{prob}$}        & 0.962      & 0.989      & 0.909      & 0.987 \\\hline
     
\end{tabular}
\caption{Magnetic exponents for the diluted O($3$) model with
$p$=0.9. Extrapolations are carried out without disregarding data.}
\label{expmag09}
\end{center}
\end{table}

\begin{table}[!ht]
\begin{center}
\begin{tabular}{c|c|c|c|c|}\cline{2-5}
& \multicolumn{4}{c|}{$\nu$} \\
\cline{1-5}
\multicolumn{1}{|c|}{$L$} & \multicolumn{1}{c|}{$\partial_\beta g_4^V$}      
    & \multicolumn{1}{c|}{$\partial_\beta \xi^V$} 
    & \multicolumn{1}{c|}{$\partial_\beta g_4^T$}        
    & \multicolumn{1}{c|}{$\partial_\beta \xi^T$}\\\hline\hline

\multicolumn{1}{|c|}{8}  & 0.7319(49) & 0.7443(24) & 0.7128(83) & 0.7709(29)\\\cline{1-5}
\multicolumn{1}{|c|}{12} & 0.7381(53) & 0.7411(25) & 0.7267(86) & 0.7514(29)\\\cline{1-5}
\multicolumn{1}{|c|}{16} & 0.7430(55) & 0.7381(26) & 0.7536(99) & 0.7426(31)\\\cline{1-5}
\multicolumn{1}{|c|}{24} & 0.7384(57) & 0.7368(28) & 0.7337(95) & 0.7395(32)\\\cline{1-5}
\multicolumn{1}{|c|}{32} & 0.7398(54) & 0.7365(29) & 0.7241(97) & 0.7345(33)\\\hline\hline
                                                
\multicolumn{1}{|c|}{$L\rightarrow\infty$ }  & 0.734(15) & 0.7318(33) & 0.728(17) & 0.7152(39)\\\cline{1-5}
\multicolumn{1}{|c|}{$\mathrm{\chi^2/d.o.f}$}& 0.134/1   & 0.168/3    & 5.468/2   & 3.156/3 \\\cline{1-5}
\multicolumn{1}{|c|}{$\mathrm{prob}$}        & 0.714     & 0.983      & 0.065     & 0.368\\\hline
    
\end{tabular}
\caption{Thermal exponents for the diluted O($3$) model with $p$=0.9. In the first and third columns we obtain bad results because not monotonally-decreasing series.}
\label{expter09}
\end{center}
\end{table}

\begin{table}[!ht]
\begin{center}
\begin{tabular}{c|c|c|c|c|}\cline{2-5}
& \multicolumn{2}{c|}{$\eta$} & \multicolumn{2}{c|}{$\eta_T$}\\
\cline{1-5}
\multicolumn{1}{|c|}{$L$}& \multicolumn{1}{c|}{$\chi$}      
    & \multicolumn{1}{c|}{\lower2pt\hbox{$M$}}
    & \multicolumn{1}{c|}{$\chi_T$}        
    & \multicolumn{1}{c|}{\lower2pt\hbox{$M_T$}}\\\hline\hline
\multicolumn{1}{|c|}{8 } & 0.0436(38) & 0.0412(41) & 1.3882(52) & 1.3879(53)\\\cline{1-5}
\multicolumn{1}{|c|}{12} & 0.0411(34) & 0.0401(36) & 1.4005(48) & 1.4007(49)\\\cline{1-5}
\multicolumn{1}{|c|}{16} & 0.0392(31) & 0.0392(34) & 1.4061(45) & 1.4073(46)\\\cline{1-5}
\multicolumn{1}{|c|}{24} & 0.0383(29) & 0.0386(31) & 1.4131(41) & 1.4136(43)\\\cline{1-5}
\multicolumn{1}{|c|}{32} & 0.0382(27) & 0.0389(29) & 1.4142(40) & 1.4149(41)\\\hline\hline

\multicolumn{1}{|c|}{$L\rightarrow\infty$}   & 0.0343(57) & 0.0370(58) & 1.4299(72) & 1.4318(76)\\\cline{1-5}
\multicolumn{1}{|c|}{$\mathrm{\chi^2/d.o.f}$}& 0.232/3    & 0.059/3    & 0.472/3    & 0.567/3   \\\cline{1-5}
\multicolumn{1}{|c|}{$\mathrm{prob}$}        & 0.972      & 0.996      & 0.925      & 0.904     \\\hline
\end{tabular}
\caption{Magnetic exponents for the diluted O($3$) model with $p$=0.7. Extrapolations are carried out without disregarding data.}
\label{expmag07}
\end{center}
\end{table}

\begin{table}[!ht]
\begin{center}
\begin{tabular}{c|c|c|c|c|}\cline{2-5}
& \multicolumn{4}{c|}{$\nu$}\\
\cline{1-5}
\multicolumn{1}{|c|}{$L$} & \multicolumn{1}{c|}{$\partial_\beta g_4^V$}      
    & \multicolumn{1}{c|}{$\partial_\beta \xi^V$} 
    & \multicolumn{1}{c|}{$\partial_\beta g_4^T$}        
    & \multicolumn{1}{c|}{$\partial_\beta \xi^T$}\\\hline\hline
\multicolumn{1}{|c|}{8}  & 0.7888(69)  & 0.7881(31) & 0.8256(143) & 0.8422(42)\\\cline{1-5}
\multicolumn{1}{|c|}{12} & 0.7810(74)  & 0.7806(33) & 0.8078(140) & 0.8067(41)\\\cline{1-5}
\multicolumn{1}{|c|}{16} & 0.7633(70)  & 0.7760(35) & 0.7739(131) & 0.7897(43)\\\cline{1-5}
\multicolumn{1}{|c|}{24} & 0.7491(66)  & 0.7628(37) & 0.7719(146) & 0.7792(47)\\\cline{1-5}
\multicolumn{1}{|c|}{32} & 0.7400(67)  & 0.7521(42) & 0.7656(178) & 0.7627(56)\\\hline\hline

\multicolumn{1}{|c|}{$L\rightarrow\infty$ }  &  0.7206(88)& 0.723(10)& 0.729(19)& 0.7255(61)\\\cline{1-5}
\multicolumn{1}{|c|}{$\mathrm{\chi^2/d.o.f}$}&  2.313/3   & 0.281/1  & 1.314/3  & 1.965/3 \\\cline{1-5}
\multicolumn{1}{|c|}{$\mathrm{prob}$}        &  0.510     & 0.596    & 0.726    & 0.580 \\\hline
\end{tabular}
\caption{Thermal exponents for the diluted O($3$) model with $p$=0.7. In the second column the fit is obtained disregarding data with $L<16$.}
\label{expter07}
\end{center}
\end{table}

\begin{table}[!ht]
\begin{center}
\begin{tabular}{c|c|c|c|c|}\cline{2-5}
& \multicolumn{2}{c|}{$\eta$} & \multicolumn{2}{c|}{$\eta_T$}\\
\cline{1-5}
\multicolumn{1}{|c|}{$L$}& \multicolumn{1}{c|}{$\chi$}      
    & \multicolumn{1}{c|}{\lower2pt\hbox{$M$}}
    & \multicolumn{1}{c|}{$\chi_T$}        
    & \multicolumn{1}{c|}{\lower2pt\hbox{$M_T$}}\\\hline\hline
\multicolumn{1}{|c|}{8 } & 0.0505(45) & 0.0461(48) & 1.3435(61) & 1.3431(62) \\\cline{1-5}
\multicolumn{1}{|c|}{12} & 0.0448(39) & 0.0439(42) & 1.3684(54) & 1.3702(56) \\\cline{1-5}
\multicolumn{1}{|c|}{16} & 0.0421(36) & 0.0417(39) & 1.3877(51) & 1.3896(52) \\\cline{1-5}
\multicolumn{1}{|c|}{24} & 0.0396(32) & 0.0406(35) & 1.4033(46) & 1.4053(48) \\\cline{1-5}
\multicolumn{1}{|c|}{32} & 0.0399(30) & 0.0414(32) & 1.4126(43) & 1.4152(45)\\\hline\hline

\multicolumn{1}{|c|}{$L\rightarrow\infty$}   & 0.0346(60) & 0.0378(46) & 1.446(12) & 1.449(12)\\\cline{1-5}
\multicolumn{1}{|c|}{$\mathrm{\chi^2/d.o.f}$}& 2.225/2    & 2.191/3    & 0.119/1   & 0.327/1  \\\cline{1-5}
\multicolumn{1}{|c|}{$\mathrm{prob}$}        & 0.329      & 0.534      & 0.730     & 0.568    \\\hline
\end{tabular}
\caption{Magnetic exponents for the diluted O($3$) model with $p$=0.5. In the third and fourth columns we have only used data with $L>12$.}
\label{expmag05}
\end{center}
\end{table}

\begin{table}[!ht]
\begin{center}
\begin{tabular}{c|c|c|c|c|}\cline{2-5}
& \multicolumn{4}{c|}{$\nu$}\\
\cline{1-5}
\multicolumn{1}{|c|}{$L$} & \multicolumn{1}{c|}{$\partial_\beta g_4^V$}      
    & \multicolumn{1}{c|}{$\partial_\beta \xi^V$} 
    & \multicolumn{1}{c|}{$\partial_\beta g_4^T$}        
    & \multicolumn{1}{c|}{$\partial_\beta \xi^T$}\\\hline\hline
\multicolumn{1}{|c|}{8}  & 0.8102(91) & 0.8357(46) & 0.9180(241) & 0.9540(72) \\\cline{1-5}
\multicolumn{1}{|c|}{12} & 0.8042(90) & 0.8322(50) & 0.8880(248) & 0.8866(71) \\\cline{1-5}
\multicolumn{1}{|c|}{16} & 0.7764(89) & 0.7862(48) & 0.8449(242) & 0.8136(64) \\\cline{1-5}
\multicolumn{1}{|c|}{24} & 0.7702(93) & 0.7778(52) & 0.8311(234) & 0.7952(66) \\\cline{1-5}
\multicolumn{1}{|c|}{32} & 0.7562(91) & 0.7779(56) & 0.7812(220) & 0.7833(70)\\\hline\hline

\multicolumn{1}{|c|}{$L\rightarrow\infty$ }  & 0.720(16) & 0.764(14)  & 0.735(28) & 0.744(17)\\\cline{1-5}
\multicolumn{1}{|c|}{$\mathrm{\chi^2/d.o.f}$}& 1.149/2   & 0.208/1    & 1.565/3   & 0.025/1 \\\cline{1-5}
\multicolumn{1}{|c|}{$\mathrm{prob}$}        & 0.563     & 0.649      & 0.667     & 0.874 \\\hline
\end{tabular}
\caption{Thermal exponents for the diluted O($3$) model with $p$=0.5. In the first column we have only used data with $L>8$ while in the second and fourth ones we have only use data with $L>12$.}
\label{expter05}
\end{center}
\end{table}

\begin{table}[!ht]
\begin{center}
\begin{tabular}{c|c|c|c|c|c|}\cline{2-5}
& \multicolumn{2}{c|}{$\eta$} & \multicolumn{2}{c|}{$\eta_T$}     \\

\cline{2-5}
    & \multicolumn{1}{c|}{$\chi$}      
    & \multicolumn{1}{c|}{\lower2pt\hbox{$M$}} 
    & \multicolumn{1}{c|}{$\chi_T$}        
    & \multicolumn{1}{c|}{\lower2pt\hbox{$M_T$}}
    \\\cline{2-5}\hline
     
\multicolumn{1}{|c|}{$\mathrm{Our\ results}$}    & 0.0390(9)& 0.0389(10)& 1.4251(13)& 1.4251(14) \\\cline{1-5}
\multicolumn{1}{|c|}{$\mathrm{\chi^2/d.o.f}$}    & 6.675/12 & 5.104/15  & 9.151/10  & 13.931/11  \\\cline{1-5}
\multicolumn{1}{|c|}{$\mathrm{prob}$}            & 0.878    & 0.991     & 0.518     & 0.237  \\\hline\hline     
\multicolumn{1}{|c|}{$\mathrm{Ref.}$~\cite{peli}}& 0.0378(6)& ---       & ---       & --- \\\cline{1-5}

\end{tabular}
\caption{Joined extrapolation with all $p$ values for the magnetic exponent $\eta$ compared with the results from Ref.~\cite{peli}.
The first three rows correspond to our $L\rightarrow\infty$ extrapolation, being $\mathrm{prob}$
the probability to find a larger value for the $\chi^2$ of the fit (it is a measure of the goodness of the fit) and $\mathrm{d.o.f.}$ the number of degrees of freedom (the total number of data minus the total number of adjustable parameters in the fitting function).
}
\label{comparationmag}
\end{center}
\end{table}

\begin{table}[!ht]
\begin{center}
\begin{tabular}{c|c|c|c|c|c|}\cline{2-5}
&  \multicolumn{4}{c|}{$\nu$} \\

\cline{2-5}
    & \multicolumn{1}{c|}{$\partial_\beta g_4^V$}      
    & \multicolumn{1}{c|}{$\partial_\beta \xi^V$} 
    & \multicolumn{1}{c|}{$\partial_\beta g_4^T$}        
    & \multicolumn{1}{c|}{$\partial_\beta \xi^T$}
    \\\cline{2-5}\hline
     
\multicolumn{1}{|c|}{$\mathrm{Our\ results}$}    & 0.7126(46)& 0.7129(31)& 0.7294(81)& 0.7089(32)\\\cline{1-5}
\multicolumn{1}{|c|}{$\mathrm{\chi^2/d.o.f}$}    & 4.831/11  & 6.606/6   & 9.009/13  & 9.609/7   \\\cline{1-5}
\multicolumn{1}{|c|}{$\mathrm{prob}$}            & 0.939     & 0.359     & 0.772     & 0.212\\\hline\hline
\multicolumn{1}{|c|}{$\mathrm{Ref.}$~\cite{peli}}& 0.7113(11)& ---       & ---       & --- \\\cline{1-5}

\end{tabular}
\caption{Joined extrapolation with all $p$ values for the thermal exponent $\nu$ compared with the results from Ref.~\cite{peli}.}
\label{comparationtherm}
\end{center}
\end{table}

\begin{table}[htbp]
\begin{center}
\begin{tabular}{c|c|c|}\cline{2-3}
    & \multicolumn{1}{c|}{$g_4^V$}        
    & \multicolumn{1}{c|}{$g_4^T$}\\\cline{2-3}\hline
     
\multicolumn{1}{|c|}{$\mathrm{Our\ results}$}     & 0.62018(6) & 0.51366(19)\\\cline{1-3}
\multicolumn{1}{|c|}{$\mathrm{\chi^2/d.o.f}$}     & 10.324/9   & 5.980/10   \\\cline{1-3}
\multicolumn{1}{|c|}{$\mathrm{prob}$}             & 0.325      & 0.817      \\\hline\hline
\multicolumn{1}{|c|}{$\mathrm{Ref.}$ \cite{peli}} & 0.6202(1)  & ---        \\\cline{1-3}
\multicolumn{1}{|c|}{$\mathrm{Ref.}$ \cite{korut}}& 0.625783   & ---        \\\cline{1-3}

\end{tabular}
\caption{Joined extrapolation to $L\rightarrow\infty$ with all $p$ values for the Binder cumulant, $g_4$, defined in Eq.~(\ref{g4cumulants}), compared with results from Refs.\cite{peli}~and~\cite{korut}.
}
\label{comparationbind}
\end{center}
\end{table}

\cleardoublepage

\end{document}